\documentstyle[12pt]{article}
\def\tdfullfigure #1 #2 #3 #4 #5 #6
  {\begin{figure}\vspace*{#3 cm}\vspace{12.3cm}
   \tdpsinput #1 #6 #5 #3
   \vspace{-4.0cm}\vspace{#4 cm}
   \caption[]{#2}\end{figure}
   \typeout{TDINPUT: FIGURE \thefigure. produced with file #1}}
\def\tdpsinput#1 #2 #3 #4 {
   \special{ps::
-1 1 scale
-90 rotate
-1700 -2342 translate
#2 #2 scale
#4 -118 mul #3 -118 mul translate
3.5 3.5 scale
      }
   \special{ps: plotfile #1 asis}
   \special{ps:: endexecute
      }}
\begin{document}
\vspace*{1mm}
{\raggedleft YCTP-N3-96\\}
\mbox{}
\vspace{1cm}
\begin{center}
{\Large\sc Topological Dependence of Universal\\
\vspace{2mm}
 Correlations in Multi-Parameter Hamiltonians}

\vspace{1.8cm}

David MITCHELL and 
Dimitri KUSNEZOV\footnote{E--mail: dimitri@nst4.physics.yale.edu} \\

\vspace{8mm}

{\em Center for Theoretical Physics, Sloan Physics Laboratory,} \\ 
    {\em Yale University, New Haven, CT 06520-8120 USA}

\vskip 1.2 cm

{\it April 1996}

\vspace{1.2cm}

\parbox{13.0cm}
{\begin{center}\large\sc ABSTRACT \end{center}
{\hspace*{0.3cm} 

Universality of correlation functions obtained in parametric random matrix 
theory is explored in a multi-parameter formalism, through the introduction of
a diffusion matrix $D_{ij}({\bf R})$, and compared to results from a 
multi-parameter chaotic model. We show that certain universal  correlation
functions in 1-d are no longer well defined by the metric distance between the
points in parameter space, due to a global topological dependence on the path
taken. By computing the density of diabolical points, which is found to
increases quadratically with the dimension of the space, we find a universal 
measure of the density of diabolical points in chaotic systems.}}
\end{center}

\vspace{3mm}

{\bf PACS numbers:}  05.40+j, 05.45+b, 24.60.-k

\newpage
%%%%%%%%%%%%%%%%%%%%%%%%%%%%%%%%%%%%%%%%%%%%%%%%%%%%%%%%%%%%%%%%%%%%%%%%%%%%%%%
\setcounter{page}{2} 
%%%%%%%%%%%%%%%%%%%%%%%%%%%%%%%%%%%%%%%%%%%%%%%%%%%%%%%%%%%%%%%%%%%%%%%%%%%%%%%

\begin{center}
{\large 1. Introduction}
\end{center}
\vspace{7mm}

The relation between classically chaotic systems and the fluctuation properties
of the corresponding quantum systems has been a focus of intense investigation
in recent years\cite{Boh91}. We have learned that the quantum counterpart of
classically chaotic systems have many properties which can be described by
random matrix theory (RMT). As a consequence, one now typically invokes random
matrix ensemble averaging to simplify the computation of properties of complex
systems. But while the understanding of properties of RMT is now quite well
established, until recently very little was known about random matrix or
chaotic Hamiltonians which depend on an external parameter $x$. The parameter
dependence can be the strength of an external field, or even the motion of slow
variables in a complex system treated in the Born-Oppenheimer
approximation\cite{Weid}. The types of questions one is interested in are how
the properties of such systems are correlated in that parameter. In the past
few years it has become apparent that a more general class of random matrix
results exist for parametric Hamiltonians. In a series of papers, Szafer-Simons
and Altshuler \cite{SLA} computed the density-density and level velocity
correlators in a parametric random matrix model, and demonstrated that it
described the behavior of very different chaotic and disordered systems. By
introducing a specific scaling of the parameter $x$, correlation functions were
found to fall upon universal curves, independent of the underlying properties
of the studied systems. Since then, there has been extensive studies on
universality in other systems, and the class of observables has been extended
to include wavefunctions and distributions of matrix elements\cite{waveold}.
Several further studies were based on Ref. \cite{waveold}, including a check
that the predictions for wavefunctions hold in chaotic systems\cite{after}, a
general theory for scaling\cite{caio}, and the introduction and verification of
a wide class of universal correlation functions and  distributions\cite{dave}.
However, all this work has been relegated to one parameter models. The purpose
of this paper is to extend the previous results to two or more parameters, and
to examine the global topological effects on universal functions due to the
multi--dimensional parameter space. We point out that the universality of
parametric correlations which emerges on short distance scales can be different
from 1-parameter model predictions.

\vspace{1cm}
\begin{center}
{\large 2. Geometric Phase for GOE Matrices}
\end{center}
\vspace{7mm}

We consider multiparameter Hamiltonians which exhibit chaos in some region
of parameter space, such that the fluctuation properties of the quantum
Hamiltonian in that regime are described by random matrix theory. We will
limit the discussion to the gaussian orthogonal ensemble (GOE), or systems
with time-reversal symmetry. As a result, the Hamiltonian is a real
symmetric matrix, with a non-degenerate eigenvalue spectrum. Consider now
the general adiabatic variation of the Hamiltonian $H({\bf R}(t))$, where
the adiabatic eigenstates are defined as 
\begin{equation}
H({\bf R}(t))\psi _n({\bf R}(t))=E({\bf R}(t))\psi _n({\bf R}(t)).
\end{equation}
For a wavefunction adiabatically transported along some path ${\bf R}$, the
wavefunction acquires both a dynamical and a geometric phase, $\gamma$\cite
{berry}: 
\begin{equation}
\Psi (t)=e^{-(i/\hbar )\int_0^tdt^{\prime }E_n({\bf R}(t^{\prime
}))}e^{i\gamma _n(t)}\psi _n({\bf R}(t))
\end{equation}
We are only concerned here with the geometric component of the acquired
phase, $\gamma _n$, which contains information about the topology of the
parameter space and about the presence of diabolical points. For the
parallel transport of the adiabatic states $\psi _n$ around a closed circuit 
$C$ in parameter space, we have 
\begin{equation}
\langle \psi _{n,final}\mid \psi _{n,initial}\rangle =e^{i\gamma _n(C)}
\end{equation}
For real symmetric matrices, the acquired geometric phase for a closed loop
is $e^{i\gamma (C)}=1$ if the path does not enclose a degeneracy (diabolical
point), and $e^{i\gamma (C)}=-1$ if the path does enclose a degeneracy\cite
{BW}. The path can be in any region of parameter space, but if the chaotic
properties of the Hamiltonian go away for a particular range of parameter,
one might encounter degeneracies and further complications such as
non-abelian gauge potentials can develop. We will not consider this here,
and hence restrict our discussion to paths in the classically chaotic
regions of the model we study, and equivalently, Hamiltonians chosen from
the GOE ensemble.

\vspace{1cm}
\begin{center}
{\large 3. Two Parameter Models}
\end{center}
\vspace{7mm}

\noindent{\sc 3.1 Random Matrix Hamiltonians}
\vspace{7mm}

We first consider a two parameter realization of the GOE ensemble. A
convenient Hamiltonian is \cite{wilk1}: 
\begin{equation}  \label{eq:twop}
H({\bf R})= H(x,y)= \frac{1}{\sqrt2}\left[ H_1\cos {x} + H_2\sin {x}+
H_3\cos {y} + H_4\sin {y}\right]\;,
\end{equation}
where $H_\alpha$ are independent, $N\times N$, GOE matrices: 
\begin{eqnarray}
\overline{H_\alpha^{ij}}&=&0 \\
\overline{H_\alpha^{ij}H_\beta^{kl}}&=&\frac{a^2}{2}\delta _{\alpha\beta}\,
(\delta_{ik}\delta _{jl}+\delta _{il}\delta_{jk}).  \nonumber
\end{eqnarray}
The parameter $a$ is related to the average level spacing $\Delta $ via 
$a=\Delta \sqrt{N}/\pi $. $H(x,y)$ is clearly GOE at every point $(x,y)$, and
has only two relevant moments: 
\begin{eqnarray}
\overline{H_{ij}(x,y)} &=&0  \nonumber \\
\overline{H_{ij}(x,y)H_{kl}(x^{\prime },y^{\prime })} &=&\frac{a^2}
2F(x-x^{\prime },y-y^{\prime })(\delta _{ik}\delta _{jl}+\delta
_{il}\delta_{jk})\;,
\end{eqnarray}
with 
\begin{equation}
F(x,y)=\frac{1}{2}\left[\cos (x)+\cos (y)\right]\cong 1-\frac{1}{4}
(x^2+y^2)+\,\cdots,\qquad\qquad F(0)=1.
\end{equation}
When only one parameter is used Eq. (\ref{eq:twop}), for example keeping $y$
fixed, we have the equivalent process: 
\begin{equation}
\overline{H_{ij}(x)H_{kl}(x^{\prime })}=\frac{a^2}2F(x-x^{\prime })
(\delta_{ik}\delta _{jl}+\delta _{il}\delta_{jk})\;,
\end{equation}
where 
\begin{equation}
F(x)=\frac{1}{2}(1+\cos (x))=\cos ^2(\frac x2)\cong 1-\frac{1}{4}
x^2+\,\cdots. 
\end{equation}

\vspace{7mm}
\noindent{\sc 3.2 A Chaotic Hamiltonian}
\vspace{7mm}

We will verify the topological effects we present here in a realistic model.
We use the Interacting Boson Model (IBM)\cite{fi}, which is built from
scalar ($s^{\dagger }$) and quadrupole $(d_\mu ^{\dagger },\mu =\pm 2,\pm
1,0)$ bosons, which carry angular momentum $L=0$ and $2$, respectively. This
model describes the low energy collective excitations of nuclei with even
numbers of protons and neutrons, the bosons representing paired nucleons. In
the consistent-Q form\cite{rick}, the Hamiltonian has only two relevant
parameters $\eta $ and $\chi $, and in spherical tensor notation, is given
by: 
\begin{equation}
H_{ibm}(\eta ,\chi )=E_0+\eta \hat n_d+\frac{1-\eta }{N_b}{\bf \hat Q}^\chi
\cdot {\bf \hat Q}^\chi +c_3{\bf \hat L}\cdot {\bf \hat L},  \label{eq:qham}
\end{equation}
Here $\hat n_d=d^{\dagger }\cdot \tilde d$ is the $d-$boson number operator, 
$\hat L_\mu =\sqrt{10}[d^{\dagger }\times \tilde d]_\mu ^{(1)}$ is the
angular momentum, $N_b$ is the boson number defined as half the number of
valence nucleons, and $\hat Q_\mu ^\chi =d_\mu ^{\dagger }s+s^{\dagger
}\tilde d_\mu +\chi [d^{\dagger }\times \tilde d]_\mu ^{(2)}$ is the
quadrupole operator. Because angular momentum is a good quantum number, the
parameter $c_3$ adds only an overall constant to $H_{ibm}$, and is
unimportant. The typical physical range of the parameters is $-\sqrt{7}
/2\leq \chi \leq 0$ and $0\leq \eta \leq 1$. The chaotic parameter range of
this model has been mapped out in detail\cite{niall}, and we consider the
quantum properties for this classically chaotic regime. The Hamiltonian is
diagonalized in the vibrational (or $U(5)$) basis, with $N_b=25$. The
universality of one-parameter random matrix theory predictions for
wavefunctions and distributions has been established for this model and is
extensively discussed in Ref. \cite{dave}.

\vspace{7mm}
\noindent{\sc 3.2 Topological Considerations}
\vspace{7mm}

The topological properties of the parameter space can be readily taken into
account using a symplectic 2-form $\sigma _{ij}$ and a  metric tensor $g_{ij}$.
The general Riemannian structure of the manifold of quantum states have been
defined in Ref. \cite{Provost}. In the adiabatic basis $ \mid \psi _n({\bf
R})\rangle $ of $H({\bf R})$, these are determined by the real and imaginary
parts of the quantum geometric tensor\cite{berry,Provost}: 
\begin{equation}
T_{ij}=\left\langle \nabla _i\psi _n\right| \left( 1-\left| \psi
_n\left\rangle {}\right\langle \psi _n\right| \right) \left| \nabla _j\psi
_n\right\rangle =g_{ij}+\frac i2\sigma _{ij}.
\end{equation}
The antisymmetric tensor $\sigma _{ij}$ is related to the element of area of
parametric loops we study below, and the metric $g_{ij}$, the measure of
distance: 
\begin{equation}
d\sigma =\sigma _{ij}dR_i\wedge dR_j,\qquad ds^2=g_{ij}dR_idR_j.
\end{equation}
It is hence possible to incorporate the topological properties of the
parameter space into a very general formalism. However, as the random matrix
model we consider here, in the universal regime, is rather simple, we forgo
a covariant construction in this article.

\vspace{7mm}
\noindent{\sc 3.4 Universal Scaling for Multiparameter Theories}
\vspace{7mm}

Scaling of the parameters is the crucial element in obtaining universality
in 1-parameter theories. We need to extend this to multi-parameter models.
Consider the short distance diffusion of the adiabatic energies of a
Hamiltonian $H({\bf R})$. From perturbation theory, following the arguments
of Dyson\cite{waveold,dyson,caio} we have to first order: 
\begin{equation}
E_n({\bf R^{\prime }})=E_n({\bf R})+H_{nn}({\bf R^{\prime }})-H_{nn}({\bf R}
)+\cdots 
\end{equation}
It is convenient to rescale the energies by the mean level spacing $
E_n\rightarrow E_n/\Delta $, so that : 
\begin{eqnarray}
\overline{(\delta E_n)^2} &=&\overline{(E_n({\bf R^{\prime }})-E_n({\bf R}
))^2}\cong \frac 1{\Delta ^2}\overline{(H_{nn}({\bf R^{\prime }})-H_{nn}(
{\bf R}))^2} \\
&=&\frac{2a^2}{\Delta ^2}(1-F({\bf R^{\prime },R}))\cong (R_i^{\prime
}-R_i)(R_j^{\prime }-R_j)\left. \left\{ \frac{2N}{\pi ^2}\frac{\partial ^2F}{
\partial R_i^{\prime }\partial R_j^{\prime }}\right| _{{\bf R^{\prime }=R}
}\right\}   \nonumber \\
&=&{\bf (R^{\prime }-R)\cdot D(R)\cdot (R^{\prime }-R)}.  \label{eq:diffu}
\end{eqnarray}
Eq. (\ref{eq:diffu}) demonstrates that on short distance scales, the
evolution of the adiabatic energies of the Hamiltonian resemble a diffusion
process, characterized by the {\sl diffusion matrix} $D_{ij}({\bf R})$. From
the above definition, $D_{ij}({\bf R})$ can be related to the
autocorrelation of the Hamiltonian or the local curvature properties of the
energy surfaces: 
\begin{equation}
D_{ij}({\bf R})=\overline{\nabla _iE_n({\bf R})\cdot \nabla _jE_n({\bf R})}
=\left. \frac{2N}{\pi ^2}\frac{\partial ^2F({\bf R^{\prime },R})}{\partial
R_i^{\prime }\partial R_j^{\prime }}\right| _{{\bf R^{\prime }=R}}.
\label{eq:defdif}
\end{equation}
To obtain universality in correlation functions that depend on the
parameters, we must rescale the parameters ${\bf R}$. In analogy to the
1-dimensional situation, we define: 
\begin{equation}
\widetilde{{\bf R}}\equiv {\bf D}^{1/2}\cdot {\bf R}
\end{equation}
Here ${\bf D^{1/2}}$ is the {\sl square root} of the diffusion matrix ${\bf D
}$ ( ${\bf D=\left[ D^{1/2}\right] ^TD^{1/2}}$ ), a model-dependent
quantity. Because $\overline{(\delta E_n)^2}$ is positive definite, it
follows that the diffusion matrix $D_{ij}({\bf R})$ must be a positive
definite matrix. As a consequence, its square root, ${\bf D^{1/2}}$, is
always well defined. Specifically, it is an upper triangular matrix, which
can be constructed using the Cholesky factorization\cite{cholesky}. In terms
of $\widetilde{{\bf R}}$, the energy level diffusion is now parameter free: 
\begin{equation}
\overline{(\delta E_n)^2}=\widetilde{{\bf R}}^2
\end{equation}
In analogy to the 1-dimensional scaling, we now expect all model dependence
to be removed when we compute correlation functions in terms of $\widetilde{
{\bf R}}$. These are in particular the dimension of the Hilbert space $N$
and the short distance behavior of the autocorrelation function $
F(x)=1-cx^2+...$ characterized by the coefficient(s) $c$. The rescaling by $
{\bf D^{1/2}}$ removes all this dependence, resulting in parameter free, 
{\sl universal} results. We only remark here in passing that there is a more
general class of processes for which the diffusion is not smooth\cite{caio},
characterized by the short distance behavior of $F(x)$: $F(x)=1-cx^\alpha
+\cdots $, with $0\leq \alpha <2$. In this case, one can proceed as
discussed above, but one must define the derivatives of $F$ in Eq. (\ref
{eq:defdif}) as fractional derivatives\cite{frac,caio}.

In general, there is dependence of the diffusion matrix on its position in
parameter space, clearly seen in Eq. (\ref{eq:defdif}). A general parametric
Hamiltonian $H({\bf R})$ will have an autocorrelation $F({\bf R^{\prime },R})
$ which is not translationally invariant, as is the case for the IBM\
Hamiltonian, so that $D$ is parameter dependent. By construction, our random
matrix Hamiltonian is translationally invariant, so that $F({\bf R^{\prime
},R})=F({\bf R^{\prime }-R})$, and as a result $D_{ij}$ is independent of $
{\bf R}$. We find for the random matrix model (Eq. \ref{eq:twop}): 
\begin{equation}
D_{ij}({\bf R})=D_{ij}=D\delta _{ij}=\frac{N}{\pi ^2}\delta _{ij}
\label{eq:thisD}
\end{equation}
In 1-parameter models, $D_{ij}\rightarrow D_{xx}=C(0)$, the scaling
introduced in Refs. \cite{SLA}. Note that our random matrix model is
isotropic, which is certainly not the case for the IBM Hamiltonian, which
contains bilinear parameter dependence of the form $\eta \chi $. In that
case we compute the components of $D_{ij}(\eta ,\chi )$ for the energy surfaces
$E_n(\eta,\chi)$ and rescale accordingly.

Consider now the path taken from $0\rightarrow $ $\widetilde{{\bf R}}.$ Each
of the paths we follow between these points can develop a different
geometric phase $\gamma _n($ $\widetilde{{\bf R}}).$ Equivalently, we can
consider closed circuits which begin and end at a given point, denoted $
\widetilde{{\bf R}}=0.$ Using the two-parameter model in Eq. (\ref{eq:twop}),
where ${\bf R}=(x,y)$, we define such a path C as:
\begin{equation}\label{eq:path}
(0,0)\rightarrow (X_0,0)\rightarrow (X_0,Y_0)\rightarrow (0,Y_0)\rightarrow
(0,0)
\end{equation}
Upon traversing this path,  the wavefunction will develop a phase according
to 
\begin{equation}
\mid \psi _n\rangle \rightarrow e^{i\gamma _n(C)}\mid \psi _n\rangle =\pm
\mid \psi _n\rangle .
\end{equation}
From Eq. (\ref{eq:thisD}), the scaling along the circuit C in the $x$ and $y$
directions is: 
\begin{equation}\label{eq:scala}
\widetilde{R_i}=\sqrt{\frac{N}{\pi ^2}}\delta _{ij}R_j,\qquad \qquad 
\widetilde{x}=\sqrt{\frac{N}{\pi ^2}}x,\qquad \widetilde{y}=\sqrt{\frac{N}{
\pi ^2}}y.
\end{equation}
The relevant quantity is the area of the loop, $A=\Delta x\Delta y=X_0Y_0$.
This quantity can be made more useful for comparisons to parametric areas in
other Hamiltonians by defining the scaled area:  
\begin{equation}\label{eq:scal}
\widetilde{A}=\Delta \widetilde{x}\Delta
\widetilde{y}=\sqrt{D_{xx}D_{yy}}X_0Y_0.
\end{equation}
As a check of our results, we plot in Fig. 1 the ratio of the computed value 
of $\widetilde{A}$ to the predicted result in Eqs.
(\ref{eq:scala}),(\ref{eq:scal}), using the Hamiltonian of Eq. (4), for varying 
sizes of the parametric loop. We take 100 points around the loop $C$, and
compute $D_{ij}(x,y)$ by averaging over the middle third of the energy surfaces
$E_n(x,y)$. The agreement is quite good. (There is a slight systematic shift of
the results which seems to be due to the method used to unfold the energy
spectrum.)
%%%%%%%%%%%%%%%%%%%%%%%%%%%%%%%%%%%%%%%%%%%%%%%%%%%%%%%%%%%%%%%%%%%%%%%%%%%%5
%%%%%%%%%%%%               FIG  1                         %%%%%%%%%%%%%%%%%%5
%%%%%%%%%%%%%%%%%%%%%%%%%%%%%%%%%%%%%%%%%%%%%%%%%%%%%%%%%%%%%%%%%%%%%%%%%%%%5
\tdfullfigure fig1_berry.ps {Comparison of computed scaled area,
$\widetilde{A}=\sqrt{D_{xx}D_{yy}}X_0Y_0$, of the parametric loop $C$, for  the
Hamiltonian of Eq. (4), to the theoretical result from Eqs. (22)-(23). The
dashed line is the expected result. }  2. 7 -1. 1.
%%%%%%%%%%%%%%%%%%%%%%%%%%%%%%%%%%%%%%%%%%%%%%%%%%%%%%%%%%%%%%%%%%%%%%%%%%%%5

The regime of universality is roughly defined by the scale \cite{waveold} 
\begin{equation}
\left| \widetilde{{\bf R}}\right| \stackrel{<}{\sim }1.  \label{eq:range}
\end{equation}
Analogously, the regime of universal behavior for topological phase 
accumulation is 
\begin{equation}
\widetilde{A}\stackrel{<}{\sim }1.  \label{eq:arange}
\end{equation}
We will see in the
next section, that the area of the loop does not have to be large
before one finds significant effects, and saturation of these effects already
occur within the universal regime.

Our previous work in Refs. \cite{dave,waveold} examined the universal
statistical decorrelations of wavefunction related observables in a one 
parameter formalism. In Fig. 2, a typical universal result is shown for the 
distribution of  wavefunction overlaps,
$P(u;\widetilde{{\bf R}}),$ with  $ u=\langle \psi _n(\widetilde{{\bf R}})\mid
\psi _n(0)\rangle,$ for a 1-parameter path. (The vertical axis in all plots of
$P(u;\widetilde{{\bf R}})$ is rescaled to place the maximum near unity.) The
solid histogram is the result from random matrix theory, and the dashed
histogram is the result from the IBM, in particular $J^\pi =10^{+}$ states
\cite{dave}. Using $N_b=25$, this corresponds to a dimension of 211 states, of
which only the middle third of the eigenstates are used. The distribution
begins as a delta function at $\widetilde{{\bf R}}=0$, and develops into a
Porter-Thomas distribution for $\widetilde{{\bf R}}\gg 1$. The Porter-Thomas
limit is non-universal in the sense that the distribution depends explicitly on
the dimension $N$ of the space. A one--dimensional path in the 2--dimensional
parameter space (Eq. 9) has the same universal behavior when plotted as a
function of the scaled distance ${\widetilde{{\bf R}}}={\bf D} ^{\frac 12}\cdot
{\bf R}$. However, if we now allow paths in two or more dimensions, the shape
of this distribution depends on {\sl how} one reaches the position 
$\widetilde{{\bf R}}.$ 
%%%%%%%%%%%%%%%%%%%%%%%%%%%%%%%%%%%%%%%%%%%%%%%%%%%%%%%%%%%%%%%%%%%%%%%%%%%%5
%%%%%%%%%%%%               FIG  2                         %%%%%%%%%%%%%%%%%%5
%%%%%%%%%%%%%%%%%%%%%%%%%%%%%%%%%%%%%%%%%%%%%%%%%%%%%%%%%%%%%%%%%%%%%%%%%%%%5
\tdfullfigure fig2_berry.ps {One-parameter universality of the distribution
of matrix elements $P(u;{\widetilde{\bf R}})$, where  $u=\langle \psi
_n(\widetilde{{\bf R}})\mid \psi _n(0)\rangle $. The solid histogram is random
matrix prediction while the dotted histogram is  the observed result from the
Interacting Boson Model\cite{dave}.  }  2. 7 -1. 1.
%%%%%%%%%%%%%%%%%%%%%%%%%%%%%%%%%%%%%%%%%%%%%%%%%%%%%%%%%%%%%%%%%%%%%%%%%%%%5

Consider the same distributions when we traverse the
path of Eq. (\ref{eq:path}) of area $\widetilde{A}$. When we transport $N$
eigenstates around the loop, a certain number of them will develop a phase
$\gamma _n(C)=\pi $, which we denote ${\cal N}(\pi )$. The remaining ${\cal
N}(0)$ states do not acquire a phase, where $N={\cal N}(0)+ {\cal N}(\pi )$.
The fraction of states which acquire a Berry's phase is defined as:   
\begin{equation}
f=\frac{{\cal N}(\pi )}{{\cal N}(0)+{\cal N}(\pi )}.
\end{equation}
As mentioned above, because a non-zero value for ${\cal N}(\pi)$ indicates that
paths followed by certain eigenstates enclose a diabolical point, this measure 
is clearly related to the number of diabolical points enclosed by the path of 
area $A$. In Fig. 3,  the wavefunction overlap distribution function, 
$P(u;\widetilde{{\bf R}})$, is shown at various points along 
the square circuit $C$. The solid histogram
corresponds to the random matrix predictions, and the dots to the results from
the IBM. Initially, the distribution is a delta function centered at
$(x,y)=(0,0)$ . As the separation increases, the decorrelation is similar to
the results of Fig. 2. But as the circuit returns to the origin, the
distribution bifurcates into two distinct distributions, corresponding to the
existence of the Berry's phase of $\pm 1.$ As the circuit is closed, the
distribution does not go back to its original form, but is now described by two
delta functions located at $\pm 1$ of equal magnitudes (within numerical
fluctuations). The fraction $f$ is at its saturation value of 50\% for this
loop, as seen in Fig. 3(f). Hence, the short distance universality that occurs
in multiparameter systems is not simply a function of the metric distance
between the two points $0$ and $\widetilde{\bf R}$. Given only the distance
$\widetilde{\bf R}$, the distribution $P(u;\widetilde{{\bf R}})$ is not well
defined, since  one might be close to either Fig. 3(a) or 3(f), which represent
the same point. However, both are universal results when one follows a path of
the same scaled area. In this case, the universal function $P(u)$ must also
identify the area subtended by the path. Figure 4 shows a similar collection 
of distributions in the case of a smaller parametric square. As can be seen
here, only a small fraction of states subtend a diabolical point. It is clear
that quantities which are sensitive to the phase of the wavefunctions will
generally be modified in multi--parameter theories.
%%%%%%%%%%%%%%%%%%%%%%%%%%%%%%%%%%%%%%%%%%%%%%%%%%%%%%%%%%%%%%%%%%%%%%%%%%%%5
%%%%%%%%%%%%               FIG  3                         %%%%%%%%%%%%%%%%%%5
%%%%%%%%%%%%%%%%%%%%%%%%%%%%%%%%%%%%%%%%%%%%%%%%%%%%%%%%%%%%%%%%%%%%%%%%%%%%5
\tdfullfigure fig3_berry.ps {Modification of 1-parameter universality for a 2
parameter square of side $X_0=Y_0=0.32$, with N=200. The distribution 
$P(u;\widetilde{\bf R})$ is shown at the  points (X,Y) =  (a) $(0,0)$, (b)
$(0.16,0)$, (c) $(0.32,0.32)$, (d) $(0,0.32)$, (e) $(0,0.16)$, (f) $(0,0)$
(after traversal of loop). Both (f) and (a) represent the same point, showing
that universality is path dependent. The dotted histogram corresponds to
results from the IBM, on a parametric loop of similar scaled area. }  2. 7 -1.
1.
%%%%%%%%%%%%%%%%%%%%%%%%%%%%%%%%%%%%%%%%%%%%%%%%%%%%%%%%%%%%%%%%%%%%%%%%%%%%5
%%%%%%%%%%%%%%%%%%%%%%%%%%%%%%%%%%%%%%%%%%%%%%%%%%%%%%%%%%%%%%%%%%%%%%%%%%%%5
%%%%%%%%%%%%               FIG  4                         %%%%%%%%%%%%%%%%%%5
%%%%%%%%%%%%%%%%%%%%%%%%%%%%%%%%%%%%%%%%%%%%%%%%%%%%%%%%%%%%%%%%%%%%%%%%%%%%5
\tdfullfigure fig4_berry.ps { Same as Fig. 3, but for a loop of smaller area. 
In this case, $X_0=Y_0=0.03$, and N = 200. The figures correspond to  $(X,Y)=$
( in units of $X_0$) (a) $(0,0)$, (b) $(1.0,1.0)$,  (c) $(0.8,1.0)$,(d) $(0,0)$
(after traversal of loop). The smaller loop area contains fewer diabolical
points and hence has fewer states which split in (d). The dashed histogram
corresponds to the results from the IBM, which are plotted slightly offset for
better contrast. }  2. 7 -1. 1.
%%%%%%%%%%%%%%%%%%%%%%%%%%%%%%%%%%%%%%%%%%%%%%%%%%%%%%%%%%%%%%%%%%%%%%%%%%%%5

\vspace{1cm}
\begin{center}
{\large 4. Universal Density of Diabolical Points}
\end{center}
\vspace{7mm}

The effect we have described in the previous section is due to diabolical
points which are enclosed by the parametric paths.  We now consider the density
of these diabolical points. It has been shown that a necessary condition for
the occurrence of a Berry's phase of $\gamma (C)=\pi $ for a system transported
around a closed loop $C$ is the existence of a diabolical point within the
loop\cite{BW,wilk1}.  Hence the fraction $f$  of states in Fig. 3(f) and 4(d)
that have $\gamma (C)=\pi $ should be a measure of the number of diabolical
points enclosed in the area $A$, or equivalently, the density of diabolical
points. We note that we assume that as the area increases, the occurrence of a
new diabolical point is responsible for  the phase change of only one
eigenfunction. For loops of small area, which is the situation in the universal
regime, this is a reasonable assumption.  In Fig. 5 (top) we plot the fraction
of states $f$ which enclose a diabolical point as a function of the scaled
area. There are three sets of results in the figure. The crosses are the result
of varying the size of the loop, $X_0=Y_0\in (0,0.25]$, with $N=200$, so that
the scaled area varies from 0 to 1.27. The boxes correspond to varying the
dimension of the matrices from $N=50-300$, for fixed area with $X_0=Y_0=0.18$.
Finally, the open circles correspond to parametric loops in the Interacting
Boson Model, at fixed dimension. As is clearly seen the behavior is
statistically equivalent for all the results, demonstrating universality of the
results. The general features of these results are the following. We see that
the fraction $f$ increases linearly with scaled area, and that statistical
saturation of 50\% occurs for $\widetilde{A}\sim 1$. (In the Hamiltonian (4),
the relation to the area is $\widetilde{A}=NA/\pi^2$ (see Fig. 1) while in the
IBM, $\widetilde{A}\propto NA$.) We expect saturation when there are many
diabolical points, so that statistically the probability of having a phase
change for an arbitrary wavefunction is 1/2. If we define the number of
diabolical points enclosed by the circuit $C$ as $ n(C)=Nf(C),$ we conclude
that in the non-saturation regime,
\begin{equation}
f(C) = cNA,\qquad \qquad n(C)=cN^2A,
\end{equation}
where $c$ is determined from the slope of Fig. 5 (bottom). In the
saturation regime,
\begin{equation}
f(C)=\frac{1}{2},\qquad\qquad n(C)=\frac{N}{2}. \label{eq:diab2}
\end{equation}
When the area is large, $f$ is no longer a good measure of the number of
diabolical points, so the dependence of  $n(C)$ and $f(C)$ with $N$
is no longer meaningful. In Fig. 5 (bottom), we plot the small area behavior of
the fraction $f$.  The statistical errors are computed by computing the
fraction $f$ and the scaled area $\widetilde{A}$ for a number of realizations
of the parametric Hamiltonians. The results for the chaotic system (IBM), the
open  circles, are not averaged over, but come from a single parametric loops.
The general result is that the number of diabolical points grows quadratically
with the dimension of the matrix. (Saturation effects occur for larger areas).
From the results of Fig. 5, we conclude that the fractional density of
diabolical points,  $\rho$, in terms of scaled area, is a universal constant
for chaotic systems,  given by   
%%%%%%%%%%%%%%%%%%%%%%%%%%%%%%%%%%%%%%%%%%%%%%%%%%%%%%%%%%%%%%%%%%%%%%%%%%%%5
%%%%%%%%%%%%%%%%%%%%%%%%%%%%%%%%%%%%%%%%%%%%%%%%%%%%%%%%%%%%%%%%%%%%%%%%%%%%5
%%%%%%%%%%%%               FIG  5                         %%%%%%%%%%%%%%%%%%5
%%%%%%%%%%%%%%%%%%%%%%%%%%%%%%%%%%%%%%%%%%%%%%%%%%%%%%%%%%%%%%%%%%%%%%%%%%%%5
\tdfullfigure fig5_berry.ps { Universality of the fractional number of
diabolical points $f$. (top) Fraction of wavefunctions enclosing a diabolical
point as a function of the scaled area of the loop $C$, calculated at (i) fixed
dimension $N=200$, with the size of the loop varying over the range $0\leq
X_0=Y_0\leq 0.25$ (crosses); (ii) fixed loop area with $X_0=Y_0=0.18$, and the 
dimension varied from $N=25-300$ (boxes); (iii) fixed dimension in the chaotic
region of the IBM parameter space, for several loop sizes (circles). (bottom)
Linear behavior of the small area behavior of $f$. From this we find the the
density of diabolical points increases quadratically with the dimension of the
matrix. }  2. 7 -1. 1.
%%%%%%%%%%%%%%%%%%%%%%%%%%%%%%%%%%%%%%%%%%%%%%%%%%%%%%%%%%%%%%%%%%%%%%%%%%%%5
%
\begin{equation}
\rho \equiv \frac{f(C)}{\tilde A} = 0.94\pm 0.10
\end{equation}
The quantity similar to $n(C)$ was examined recently in a very interesting
study of avoided level crossings \cite{wilk1}, using the same two--parameter
GOE model.  In that study, a similar scaling in $N$  was found for the
number of diabolical points, although their results were not universal, and
do not completely agree with the results of this study. The overall scaling
agreement is encouraging, and we believe the discrepancy is due to the
different methods used to count diabolical points. 

\vspace{1cm}
\begin{center}
{\large 5. Conclusions}
\end{center}
\vspace{7mm}

We have developed a formalism to compute universal parametric correlations in
multiparameter systems, explicitly demonstrating universality in chaotic
systems. The scaling in achieved through the square root of the diffusion
matrix $D_{ij}$. We have also found that universal results for two--point
correlation functions and distributions, which are thought to be only a
function of the separation of the points in parameter space,  can be path
dependent in more than one parameter dimension. Indeed, one can obtain distinct
results for the universal correlations by reaching the same final point by
different paths. So short distance correlations are not a simple function of
the metric distance between the points in parameter space, but are also a
function of the shape of the path taken between the points. This is true for
all quantities which are sensitive to the phase of the wavefunctions. Whether
one can actually measure interference effects in a manner analogous to measures
of Berry's phase is an interesting question to explore. Quantities which are
not sensitive to the phase, will not suffer from this ambiguity. Finally, we
have verified that the density of diabolical points grows quadratically with
the dimension of the matrix, and used that to determine a  universal measure of
the density of diabolical points in chaotic systems.

This work was supported in part by the Department of Energy Grant
DE-FG02-91ER40608.

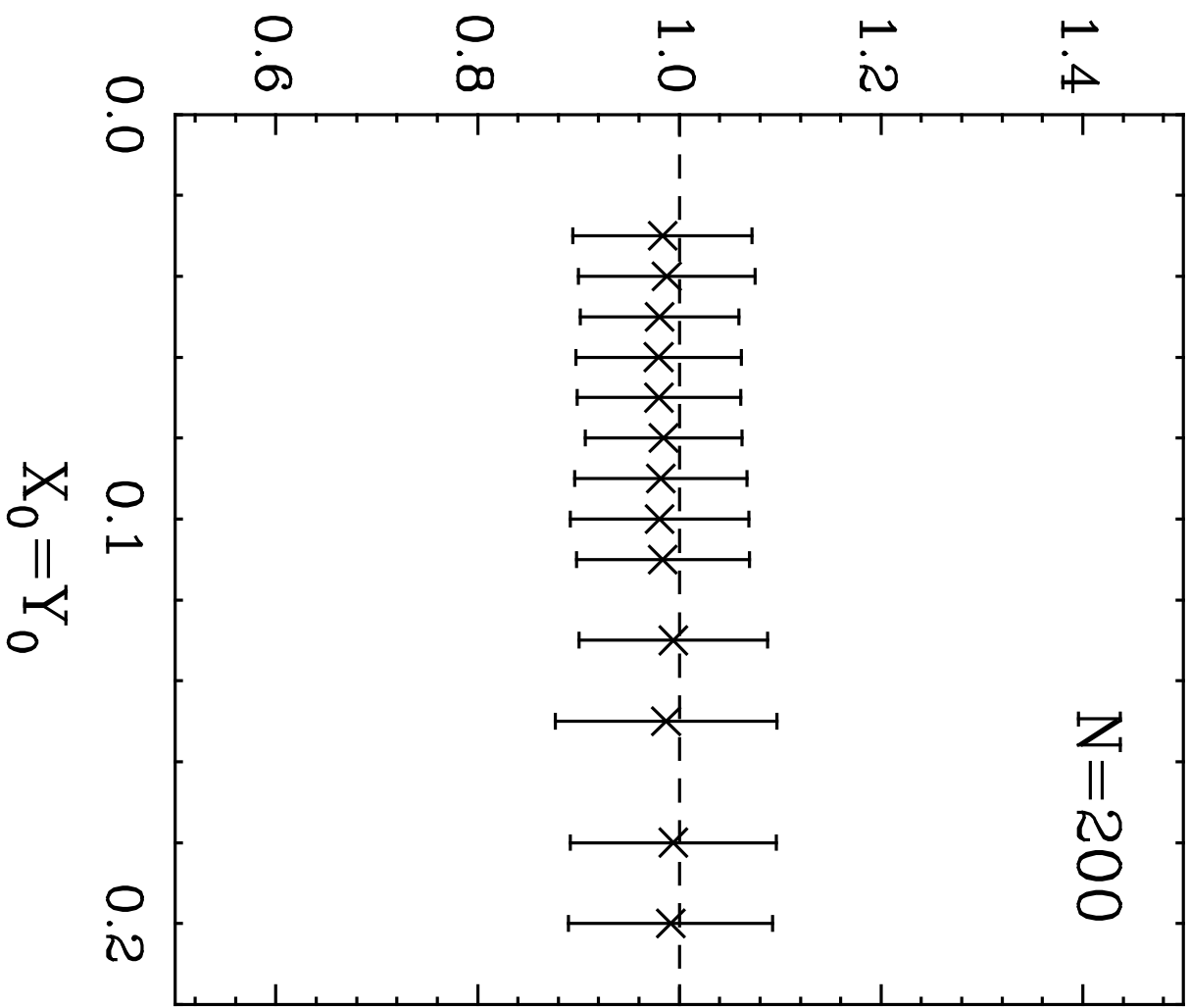
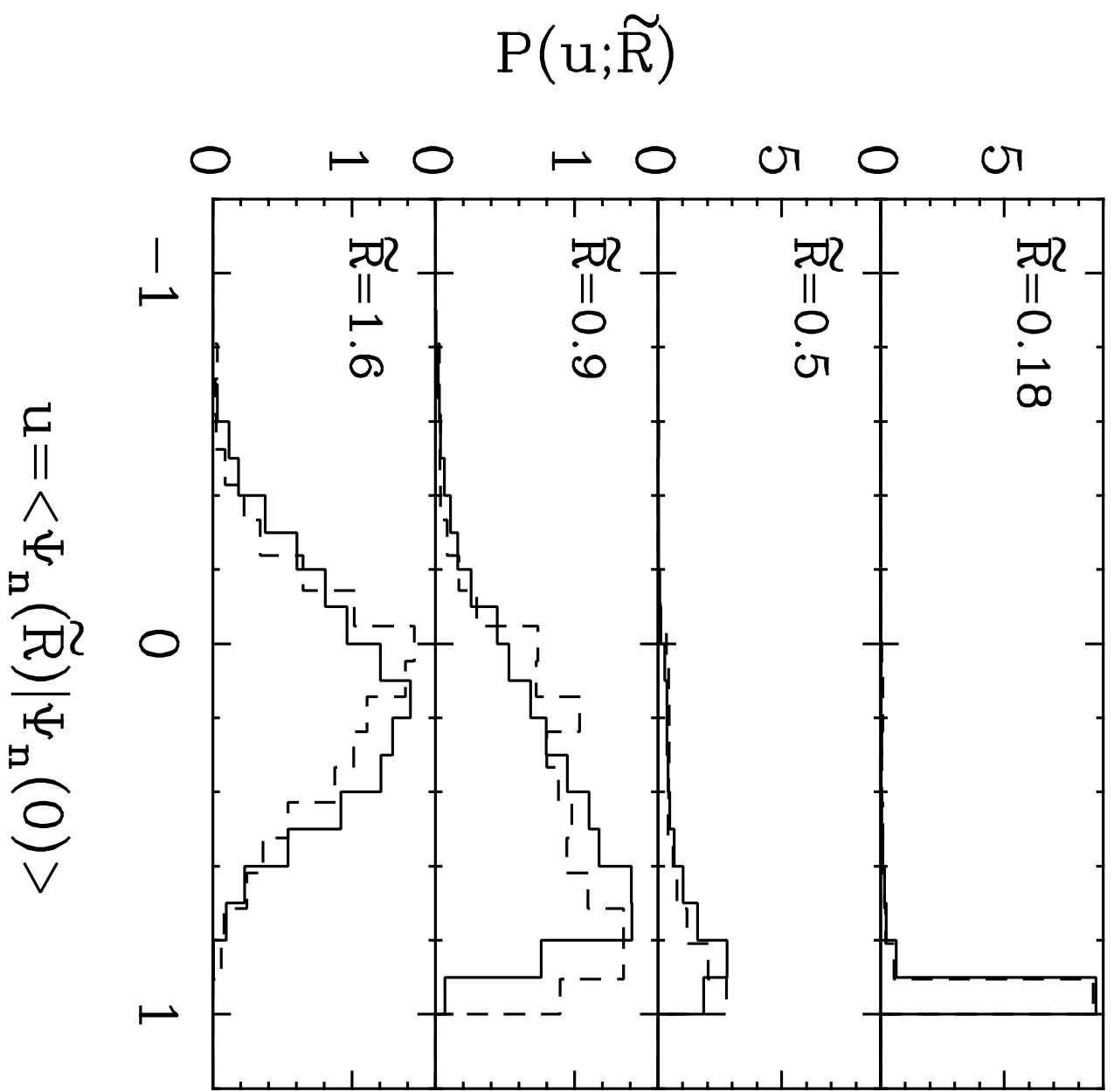
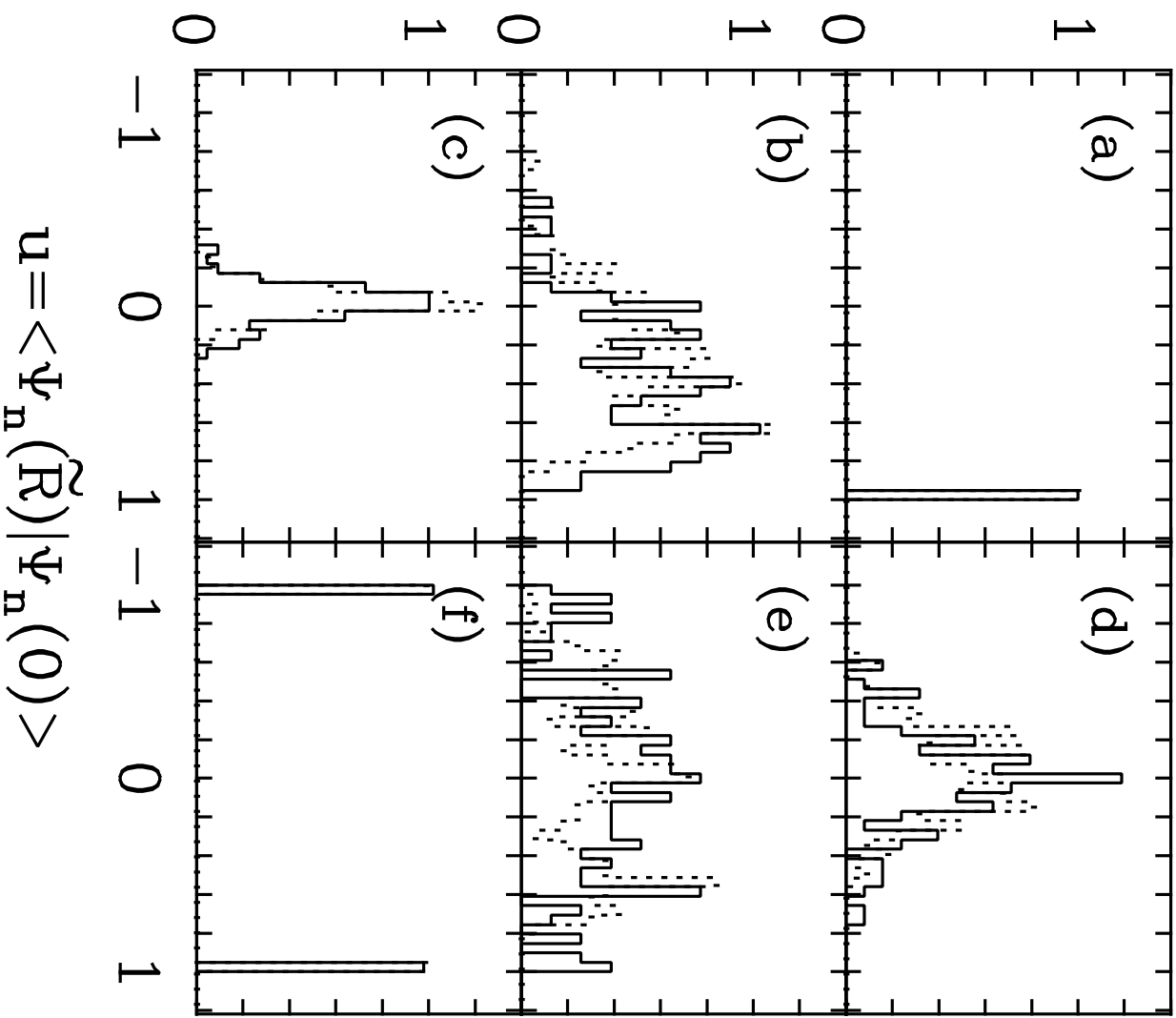
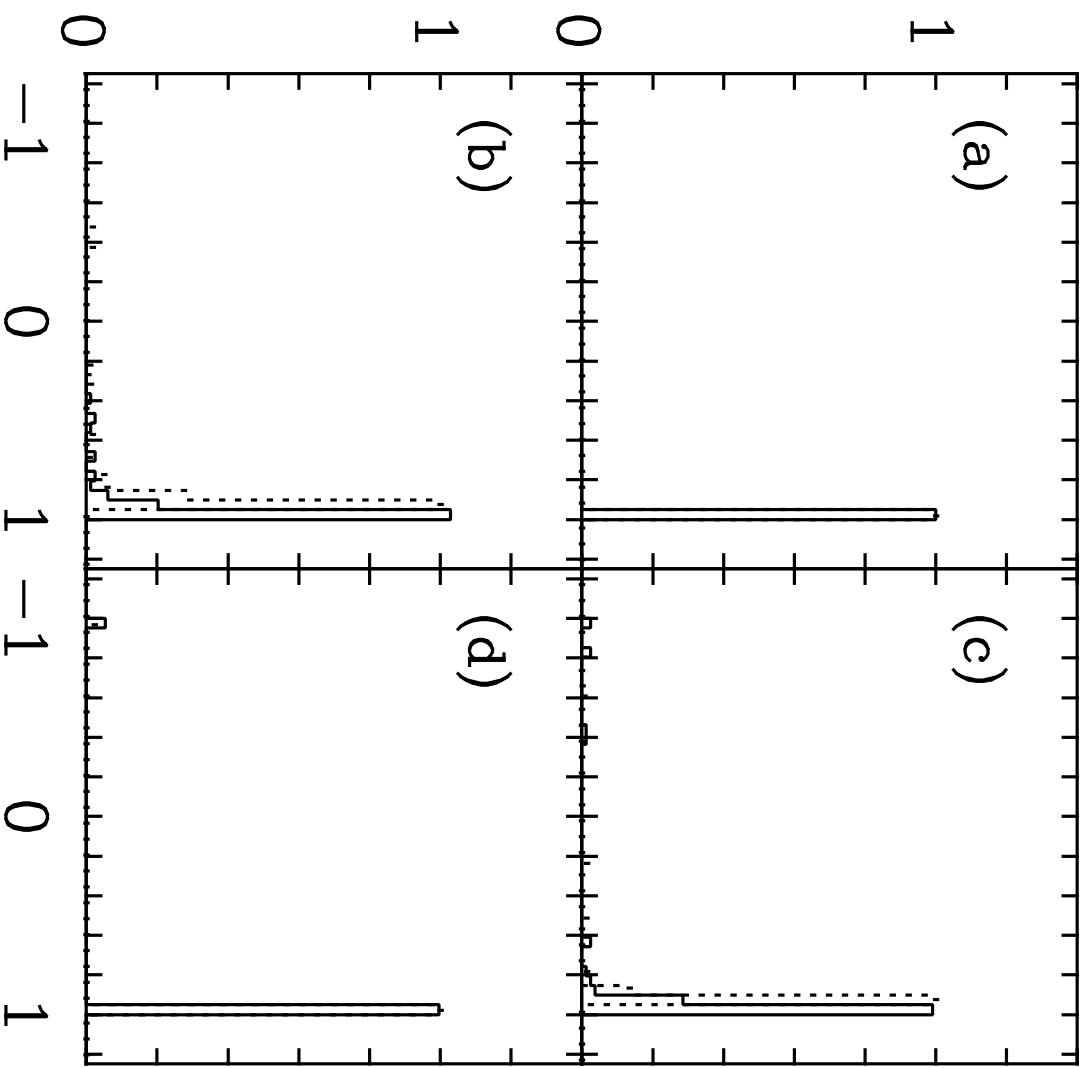
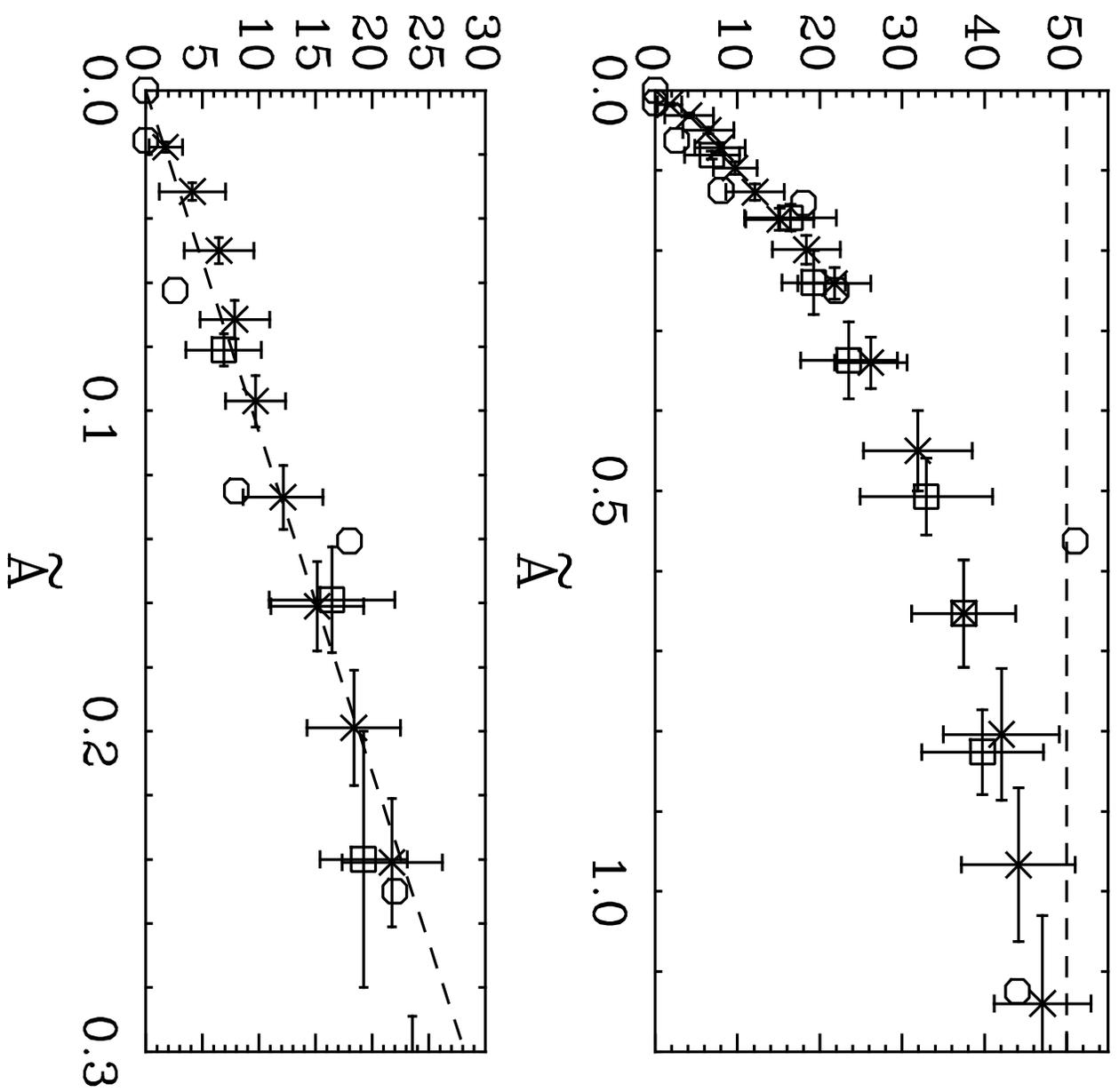
\end{document}